\documentclass[onecolumn,prd,nofootinbib,showpacs]{revtex4}

\usepackage{textcomp}
\usepackage{mathtools, ulem, graphicx}
\usepackage[export]{adjustbox}
\usepackage{sidecap}
\usepackage{verbatim}
\usepackage{feynmp}
\usepackage{slashed}
\usepackage[caption=false]{subfig}

\newcommand{\e}{\begin{equation*}\begin{aligned}}
\newcommand{\ee}{\end{aligned}\end{equation*}}

\newcommand{\en}{\begin{equation}\begin{aligned}}
\newcommand{\een}{\end{aligned} \end{equation}}

\newcommand{\pfa}[2]{\frac{\delta #1}{\delta #2}}

\newcommand{\p}{\partial}

\newcommand{\f}[2]{\frac{#1}{#2}}

\newcommand{\ra}{\rangle}
\newcommand{\la}{\langle}

\newcommand{\ma}{\mathcal}

\newcommand{\tr}{\text{tr }}

\newcommand{\Q}{\left}
\newcommand{\W}{\right}

\newcommand{\pma}{\begin{pmatrix}}
\newcommand{\epma}{\end{pmatrix}}

\newcommand{\lam}{\lambda}

\begin{document}

\title{Relationship between Fujikawa's Method and the Background Field Method for the Scale Anomaly}
\author{Chris L. Lin}
\author{Carlos R. Ord\'{o}\~{n}ez}
\affiliation{Department of Physics, University of Houston, Houston, TX 77204-5005}

\date{\today}
\email{cllin@uh.edu}
\email{cordonez@central.uh.edu}

\begin{abstract}
We show the equivalence between Fujikawa's method for calculating the scale anomaly and the diagrammatic approach to calculating the effective potential via the background field method, for an $O(N)$ symmetric scalar field theory. Fujikawa's method leads to a sum of terms, each one superficially in one-to-one correspondence with a vacuum diagram of the 1-loop expansion. From the viewpoint of the classical action, the anomaly results in a breakdown of the Ward identities due to a scale-dependence of the couplings, whereas in terms of the effective action, the anomaly is the result of the breakdown of Noether's theorem due to explicit symmetry breaking terms of the effective potential. 

\end{abstract}


\maketitle

\section{Introduction}

Fujikawa showed that within the path-integral formalism, all anomalies are the result of non-invariance of the measure under symmetry transformations \cite{fuji1,fuji2,fuji}. The resulting Jacobian then spoils the naive Ward identities. It is also known that the quantum effective action preserves the symmetries of the classical action, provided that the measure is invariant under the symmetry transformations \cite{sred}. Therefore there should be a relationship between Fujikawa's method and the non-invariant terms of the quantum effective action. We investigate this relationship in the context of an $O(N)$, $\lam \phi^4$ theory, by comparing, term-by-term, the Taylor expansion of the Fujikawa determinant with all diagrams in the 1-loop expansion of the quantum effective potential.\\

The reason for embarking on this comparison is that a framework for applying Fujikawa's method to nonrelativistic, classically scale-invariant systems was undertaken recently \cite{Ord,lin,vir}. While the quantum effective action is a standard tool in nonrelativistic physics (e.g., see \cite{braat,anderson}), Fujikawa's method is not. Therefore a comparison of the two approaches, without a coupling to a gravitational background as is done for the relativisitic case, might be helpful in a first approximation as a bridge between the two methods in the context of nonrelativistic physics.  \\

It is well known that for the chiral anomaly the choice of regulating function $f\Q(\f{\slashed{D}^2}{\Lambda^2}\W)$ one uses to regulate the Jacobian is arbitrary, except for a few conditions governing the behavior of $f$ and its derivatives at $0$ and $\infty$ that are quite reasonable \cite{Weinberg}. The argument of the regulating function however is not arbitrary - one must choose the gauge invariant $\slashed{D}$. The anomaly calculated in this manner is both finite and exact. \\

For the scale anomaly things aren't as clear. There is no symmetry that tells you what variable must go into the regulating function. Moreover, if one Taylor expands the anomaly as one does in the chiral case, certain terms are infinite. If one ignores those terms, then one can recover the anomaly, but it is not exact, holding only to 1-loop order. One generally chooses the quadratic part of effective action for the argument since it characterizes 1-loop effects \cite{shiz}.   \\

In this paper we attempt to explore the connection between certain terms in the effective potential when it is expanded by the number of vertices and certain terms in the Jacobian of Fujikawa's method when it is Taylor expanded, thereby clarifying the statement that putting the quadratic part of the effective action in the regulating function captures the 1-loop effects. Also, we consider $O(N)$ as opposed to a single scalar field because despite the problems of Fujikawa's method for the case of the scale anomaly compared to the chiral anomaly, such as only capturing the 1-loop result, it still retains a universal quality in that it can capture the 1-loop result for any $N$. \\ 

In the next two sections, we give a quick review of Fujikawa's method and the background field method for calculating the effective action. In the fourth section we apply Fujikawa's method to calculate the anomaly and the $\beta$ function of $N$ scalar fields interacting via an $O(N)$ symmetric $\lambda \phi^4$ potential. In the fifth section we use the background field method to write an expression for the effective potential, organized by the number of vertices, and compare this result with the Taylor expansion resulting from Fujikawa's method to derive conditions on the Fujikawa regulator for the two approaches to give the same result. Finally, in the sixth section we apply Noether's theorem to the effective action and compare it to anomalous scale-breaking of the classical action.

\section{Fujikawa's Method}

For simplicity we will demonstrate this method for a single scalar field: the generalization to multiple fields is straightforward. With a change of variables given by $\phi'(x)=\phi(x)+\epsilon \delta \phi(x)$:

\en
\int [d\phi] e^{iS[\phi]}&=\int [d\phi'] \Q |\pfa{\phi}{\phi'}\W | e^{iS[\phi(\phi')]} \\
&=\int [d\phi'] \Q |\delta^d(x-y) - \epsilon \pfa{\delta \phi'(x)}{\phi'(y)} \W |e^{iS[\phi'-\epsilon \delta \phi']}\\
&=\int [d\phi] \Q |\delta^d(x-y) - \epsilon \pfa{\delta \phi(x)}{\phi(y)} \W |e^{iS[\phi-\epsilon \delta \phi]}\\
&=\int [d\phi] e^{-\epsilon \int d^d x  \pfa{\delta \phi}{\phi} }e^{iS[\phi]}e^{-i\epsilon  \int d^dx \pfa{S}{\phi}\delta \phi}\\
&=\int [d\phi] e^{iS[\phi]} \Q( 1-\epsilon \int d^d x  \pfa{\delta \phi}{\phi} -i \epsilon  \int d^dx \pfa{S}{\phi}\delta \phi  \W).
\een

Since this holds for any volume $V$, it follows:

\en
\Q\la  \pfa{S}{\phi}\delta \phi \W\ra=i\Q\la   \Q.\pfa{\delta \phi(x)}{\phi(y)}\W|_{y=x} \W\ra.
\een

If $\phi \rightarrow \phi+ \epsilon \delta \phi$ is a symmetry transformation, then $\pfa{S}{\phi}\delta \phi=-\p_\mu j^\mu$, so that Fujikawa's method tells us that:

\en
\Q\la  \p_\mu j^\mu \W\ra=-i\Q\la   \Q.\pfa{\delta \phi(x)}{\phi(y)}\W|_{y=x} \W\ra.
\een

The transformation we're interested in are dilations for $N$ scalar fields:

\en
x'^\mu&=e^{-\rho}x^\mu \\
\phi'_i(x')&=e^{\rho}\phi_i(x),
\een 

so that the Jacobian is:

\en
J=\pfa{\delta \phi_i(x)}{\phi_j(y)}&=(1+x^\mu \p_\mu) \delta^4(x-y) I_n \\ &\equiv \theta \delta^4(x-y) I_n,
\een

where $I_n$ is the N-dimensional identity matrix and $\theta=(1+x^\mu \p_\mu)$.

\section{Background Field Method}

We briefly review some facts about the effective action. The generating functional $W[J]$ for the connected correlation functions can be expressed via the path integral as:

\en \label{1b}
e^{iW[J]}=\int [d\phi]\, e^{iS[\phi]+i\int J \phi}.
\een

The effective action is defined as the Legendre transform:

\en  \label{2b}
\Gamma[\phi_c]&=W[J(\phi_c)]-\int J(\phi_c) \phi_c, \\
\phi_c&=\f{\delta W}{\delta J}=\la \phi \ra_J.
\een

$\Gamma[\phi_c]$ obeys the classical equations of motion:

\en  \label{3b}
\f{\delta \Gamma }{\delta \phi_c}=-J,
\een

and can be expanded as:

\en  \label{4b}
\Gamma[\phi_c]&=\sum\limits_{n=0}^\infty \f{1}{n!} \int dx_1...dx_n \, G^{(n)}_{1PI}(x_1, ..., x_n) \phi_c (x_1)...\phi_c (x_n)\\
&=\int dx \,\Q(-V_{\text{eff}}(\phi_c)+\f{1}{2}Z(\phi_c)\p_\mu\phi_c \p^\mu \phi_c+...\W),
\een

which shows that $\Gamma[\phi_c]$ is the generating functional for the 1PI graphs and that the effective potential $V_{\text{eff}}$ is the negative sum of all 1PI graphs with all external lines set to 0 momentum. \\

In the background field method\footnote{For a review of the background field method see \cite{abbot}.}, we define a new generating functional $\tilde{W}[J]$:

\en  \label{5b}
e^{i\tilde{W}[J]}&=\int [d\phi]\, e^{iS[\phi+\hat{\phi}]+i\int J \phi}=\int [d\phi]\, e^{iS[\phi]+i\int J (\phi-\hat{\phi})}\\
&=e^{iW[J]} e^{-iJ \hat{\phi}}.
\een

Application of Eq. \eqref{2b} to $\tilde{W}[J]$ then gives the following relationships:

\en  \label{6b}
\tilde{W}[J]&=W[J]-J \hat{\phi} \\
\tilde{\phi}_c&=\phi_c-\hat{\phi} \\
\tilde{\Gamma}[\tilde{\phi}_c,\hat{\phi}]&=\Gamma[\tilde{\phi}_c+\hat{\phi}].
\een

Setting $\tilde{\phi}_c=0$ for the effective action then gives us the result we'll need:

\en \label{7b}
\Gamma[\hat{\phi}]=\tilde{\Gamma}[0,\hat{\phi}],
\een

which states that to calculate the effective action $\Gamma[\hat{\phi}]$ associated with the classical action $S[\hat{\phi}]$, we need only calculate the 1PI vacuum graphs associated with the classical action $S[\phi+\hat{\phi}]$, i.e. the original action shifted by a background $\hat{\phi}$. In the following section we will relabel $\phi$ in $S[\phi+\hat{\phi}]$ as $\eta$.

\section{Fujikawa Calculation}

Consider the conformally invariant Lagrangian

\en \label{1}
\ma L=\f{1}{2} \p_\mu \phi_i \p^\mu \phi_i-\f{\lambda}{4} (\phi_i \phi_i)^2,
\een

where repeated indices are summed and $i=1,2,...N$. The quadratic part of the action $S$ expanded around the constant background fields $\hat{\phi}_i$ ($\phi_i=\hat{\phi}_i+\eta_i$) is given by:

\en \label{2}
\tilde{S}_2= \f{1}{2} \sum \limits_{i,j=1}^{N} \int d^4x \, d^4y \,\f{\delta^2 S}{\delta \phi_j(x) \delta \phi_i(y)} \eta_j(x) \eta_i(y),
\een

which can be re-expressed in terms of the Lagrangian:

\en \label{3}
\tilde{S}_2= \f{1}{2} \sum \limits_{i,j=1}^{N} \int d^4x \Q( \f{\p^2 \ma L}{\p \phi_i \p \phi_j}\eta_i(x) \eta_j (x)+2\f{\p^2 \ma L}{\p \phi_i \p \p_\mu \phi_j}\eta_i(x) \p_\mu \eta_j (x)+\f{\p^2 \ma L}{\p \p_\nu \phi_i \p \p_\mu \phi_j} \p_\nu \eta_i(x) \p_\mu \eta_j (x)\W).
\een

Plugging in Eq. \eqref{1} into Eq. \eqref{3} gives:

\en \label{4}
\tilde{S}_2&= \f{1}{2} \sum \limits_{i,j=1}^{N} \int d^4x \Q(\Q[-2\lambda \hat{\phi}_i \hat{\phi}_j-\lambda(\hat{\phi}_k \hat{\phi}_k)\delta_{ij}\W]\eta_i(x)\eta_j(x)+\p_\mu\eta_i(x) \p^\mu \eta_i(x) \W)\\
&=\f{1}{2} \sum \limits_{i,j=1}^{N} \int d^4x \, \eta_i(x) \Q( B_{ij}+D_{ij} \W)  \eta_j(x) = \f{1}{2} \sum \limits_{i,j=1}^{N} \int d^4x \, \eta_i(x) M_{ij}  \eta_j(x),
\een

where                  

\en \label{5}
D_{ij}=-\delta_{ij} \p^2, \quad B_{ij}=\Q[-2\lambda \hat{\phi}_i \hat{\phi}_j-\lambda(\hat{\phi}_k\hat{\phi}_k)\delta_{ij}\W].
\een

We choose $M_{ij}$ as the argument of our regulating matrix so that:

\en \label{6}
\ma A= 
\text{tr} \Q[R\Q( \f{M}{\Lambda^2} \W)\theta\delta^4(x-y) I_n \W]. 
\Bigg |_{x=y}.
\een

Going into Fourier space:

\en \label{7}
\ma A&= 
\text{tr} \int \f{d^4 k}{(2\pi)^4} \Q[R\Q( \f{M}{\Lambda^2} \W)\theta e^{i k \cdot (x-y)} I_n \W] 
\Bigg |_{x=y}\\
&=\text{tr} \int \f{d^4 k}{(2\pi)^4} \Q[R\Q( \f{M}{\Lambda^2} \W)(1+x_\mu k_\mu)  I_n \W] \\
&=\Lambda^4 \,\text{tr} \int \f{d^4 k}{(2\pi)^4} \Q[R\Q( D+\f{B}{\Lambda^2} \W)  I_n \W],
\een

where in the second line $y$ has been set equal to $x$ and $D_{ij}=-\delta_{ij} \p^2 \rightarrow \delta_{ij} k^2$. $D_{ij}$ is even in $k^2$, therefore the $x_\mu k_\mu$ term vanishes upon integration. Since $[D,B]=0$, $R\Q( D+\f{B}{\Lambda^2} \W)$ admits a power series expansion about $D$:

\en \label{8}
\ma A&=\Lambda^4 \,\text{tr} \int \f{d^4 k}{(2\pi)^4} \Q[R\Q(D\W)+R'(D)\f{B}{\Lambda^2}+\f{1}{2!}R''(D)\Q(\f{B}{\Lambda^2}\W)^2+... \W].
\een

$D$ is diagonal, hence we can write $R^{(n)}(D)=f^{(n)}(k^2)I_n$ for some scalar function $f(k^2)$, so that Eq. \eqref{8} becomes:

\en \label{9}
\ma A&=\Lambda^4 N \int \f{d^4 k}{(2\pi)^4} f(k^2)+\Lambda^2 \Q(\text{tr}\, B\W) \int \f{d^4 k}{(2\pi)^4} f'(k^2)+\f{1}{2! }\Q(\text{tr}\, B^2\W) \int \f{d^4 k}{(2\pi)^4} f''(k^2)+... \\
&=
\Lambda^4 N \int \f{d^4 k}{(2\pi)^4} f(k^2)+\Lambda^2 \Q(\text{tr}\, B\W) \int \f{\Omega_3 dk^2}{2(2\pi)^4}k^2 f'(k^2)+\f{1}{2! }\Q(\text{tr}\, B^2\W) \int \f{\Omega_3 dk^2}{2(2\pi)^4} k^2 f''(k^2)\\
&+\sum \limits_{n=3}^\infty \f{1}{\Lambda^{(2n-4)}}\f{1}{n! }\Q(\text{tr}\, B^{n}\W) \int \f{\Omega_3 dk^2}{2(2\pi)^4} k^2 f^{(n)}(k^2),
\een 

where $\Omega_3=2\pi^2$ is the solid angle. The minimum conditions on $f(k^2)$ required to produce the anomaly are:

\en \label{10}
f(0)&=1 \\ 
f(\infty)&=0 \\
\Q[k^2 f'(k^2)\W]\big |^\infty_0&=0, 
\een

which are the same conditions for the chiral anomaly \cite{Weinberg}. However, for simplicity we will specialize to $f(k^2)=e^{-k^2}$, which satisfies Eq. \eqref{10}, but in addition has the nice property that:

\en
\int dk^2 k^2 f^{(n)}(k^2)=(-1)^n,
\een

so that plugging in this regulator into Eq. \eqref{9} gives us:

\en \label{11}
\ma A&=\sum \limits_{n=0}^\infty \f{(-1)^n}{\Lambda^{(2n-4)}}\f{1}{n! }\Q(\text{tr}\, B^{n}\W) \f{\Omega_3}{2(2\pi)^4} 
\\
&=\Lambda^4 \Q(\text{tr}\, B^0\W) \f{\Omega_3}{2(2\pi)^4}-\Lambda^2 \Q(\text{tr}\, B\W) \f{\Omega_3}{2(2\pi)^4} +\f{1}{2! }\Q(\text{tr}\, B^2\W) \f{\Omega_3}{2(2\pi)^4}+ 
\sum \limits_{n=3}^\infty \f{(-1)^n}{\Lambda^{(2n-4)}}\f{1}{n! }\Q(\text{tr}\, B^{n}\W) \f{\Omega_3}{2(2\pi)^4}.  \\
\een 

The first term in Eq. \eqref{11} is independent of the coupling $\lambda$ so it would be present even in the free theory. Since the free theory is taken to be non-anomalous, we ignore this term \cite{uwe}. The second term, proportional to $\Lambda^2$ is removed by mass renormalization: the precise meaning of this is discussed in the next section. The third term is the only remaining nonvanishing term in the $\Lambda \rightarrow \infty$ limit, and is independent of $\Lambda$. Evaluating $\Q(\text{tr}\, B^2\W)=B_{ij} B_{ji}$ by substituting in $B_{ij}$ from Eq. \eqref{5} gives:

\en \label{12}
\ma A&=\f{1}{2!}\Q[ \lambda^2(N+8) (\hat{\phi}_k\hat{\phi}_k)^2 \W] \f{\Omega_3}{2(2\pi)^4}\\
&=\f{\lambda^2 (N+8)}{32 \pi^2} (\hat{\phi}_k\hat{\phi}_k)^2 \\
&=\beta(\lambda)\f{(\hat{\phi}_k\hat{\phi}_k)^2 }{4}=\beta(\lambda)\f{\p \ma H_I}{\p \lambda},
\een

where $\beta(\lambda)=\f{\lambda^2 (N+8)}{8 \pi^2}$ and $\ma H_I$ is the interacting Hamiltonian. 

\section{Equivalence of Fujikawa With Background Field Calculation}

We now apply the background field method to the Lagrangian in Eq. \eqref{1}. We make the shift $\phi_i(x)=\hat{\phi}_i+\eta_i(x)$ so that the $O(N)$ Lagrangian becomes:

\en \label{1c}
\tilde{\ma L}=\f{1}{2} \sum \limits_{i,j=1}^{N} \int d^4x \, \eta_i(x) \Q( D_{ij}+B_{ij} \W)  \eta_j(x)+\ma L(\hat{\phi}_i,\p_\mu \hat{\phi}_i )+\ma L_T+\ma L_I. 
\een

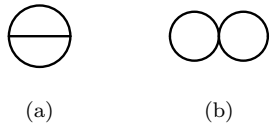
\begin{figure}%
\centering
\subfloat[]
{
\begin{fmffile}{diagram1}
\begin{fmfgraph}(40,40)
\fmfleft{i}
\fmfright{o}
\fmf{phantom,tension=5}{i,v1}
\fmf{phantom,tension=5}{v2,o}
\fmf{plain,left,tension=0.4}{v1,v2,v1}
\fmf{plain}{v1,v2}
\end{fmfgraph}
\end{fmffile}
}
\quad
\subfloat[]
{
\begin{fmffile}{diagram2}
\begin{fmfgraph}(40,40)
\fmfleft{i}
\fmfright{o}
\fmf{phantom,tension=10}{i,i1}
\fmf{phantom,tension=10}{o,o1}
\fmf{plain,left,tension=0.4}{i1,v1,i1}
\fmf{plain,right,tension=0.4}{o1,v1,o1}
\end{fmfgraph}
\end{fmffile}
}

\caption{Lowest-loop 1PI vacuum graphs with 3 and 4 vertices.}
\label{fig1}

\end{figure}

In the above expression, $\ma L(\hat{\phi}_i,\p_\mu \hat{\phi}_i )$ is the original $O(N)$ Lagrangian with the background field substituted for $\phi$. This term has no dependence on $\eta$ and contributes to the 1PI vacuum graphs at tree-level (i.e., w.r.t. the $\eta$ field this term is like a cosmological constant). $\ma L_T$ are terms that contain only one $\eta$ field: these produce tadpole diagrams which are reducible, so $\ma L_T$ can be neglected in calculation of 1PI graphs. $L_I$ are terms involving $\eta^3$ and $\eta^4$ interactions. For 1PI vacuum graphs, these interactions contribute beginning at the 2-loop level, and hence can be ignored for a 1-loop calculation (see Fig. \ref{fig1}). \\


So the Lagrangian we will use to calculate the 1PI vacuum graphs at 1-loop is:

\en \label{2c}
\tilde{\ma L}=\f{1}{2} \sum \limits_{i,j=1}^{N} \int d^4x \, \eta_i(x)  D_{ij}  \eta_j(x)+\f{1}{2} \sum \limits_{i,j=1}^{N} \int d^4x \, \eta_i(x)  B_{ij}  \eta_j(x).
\een

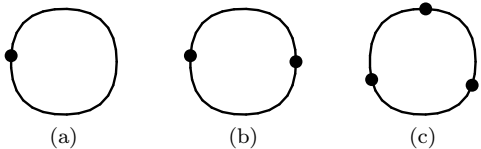
\begin{figure}%
\centering
\subfloat[]
{
\begin{fmffile}{diagram3}
\begin{fmfgraph}(40,40)
\fmfsurroundn{v}{29}\fmfdot{v15}
\fmf{plain}{v1,v2,v3,v4,v5,v6,v7,v8,v9,v10,v11,v12,v13,v14,v15,v16,v17,v18,v19,v20,v21,v22,v23,v24,v25,v26,v27,v28,v29,v1}
\end{fmfgraph}
\end{fmffile}
}
\quad
\subfloat[]
{
\begin{fmffile}{diagram4}
\begin{fmfgraph}(40,40)
\fmfsurroundn{v}{29}\fmfdot{v1,v15}
\fmf{plain}{v1,v2,v3,v4,v5,v6,v7,v8,v9,v10,v11,v12,v13,v14,v15,v16,v17,v18,v19,v20,v21,v22,v23,v24,v25,v26,v27,v28,v29,v1}
\end{fmfgraph}
\end{fmffile}
}
\quad
\subfloat[]
{
\begin{fmffile}{diagram5}
\begin{fmfgraph}(40,40)
\fmfsurroundn{v}{29}\fmfdot{v8,v17,v28}
\fmf{plain}{v1,v2,v3,v4,v5,v6,v7,v8,v9,v10,v11,v12,v13,v14,v15,v16,v17,v18,v19,v20,v21,v22,v23,v24,v25,v26,v27,v28,v29,v1}
\end{fmfgraph}
\end{fmffile}
}

\caption{1-loop 1PI vacuum graphs with 1,2, and 3 vertices.}
\label{fig2}

\end{figure}

Since the background field $\hat{\phi}_i$ (contained in $B_{ij}$ of Eq. \eqref{5}) is constant and the Lagrangian is only quadratic in $\eta$, we could sum all the 1-loop vacuum graphs at once by calculating the determinant $ D_{ij}+B_{ij}$ \cite{jackiw}. However, instead we choose as the propagator $D^{-1}_{ij}$, and treat interaction $B_{ij}$ as an interaction vertex that joins two propagators, and categorize the loops by the number of vertices $B_{ij}$ which corresponds to twice the number of background fields $\hat{\phi}$ (see Fig. \ref{fig2}). We do this to match the result of Eq. \eqref{11} from Fujikawa's method, which is an expansion in powers of $B_{ij}$. \\


The Feynman rules are straightforward. For each vertex we write $iB_{ij}$, as the $1/2$ in Eq. \eqref{2c} accounts for swapping connections of the two propagators to which each vertex connects. For each propagator we write $i D^{-1}_{ij}$, where the $1/2$ takes care of which end of the propagator connects to a vertex. An overall symmetry factor is required that depends on the number of vertices $B_{ij}$. This symmetry factor is $\f{1}{2n}$ where $n$ is the number of vertices: the 2 is due to reflection symmetry and $n$ to cyclic permutation of the vertices.       \\

For an n-vertex diagram:

\en \label{3c}
-iV^n_{\text{eff}}=\f{1}{2n}\int \f{id^4k}{(2\pi)^4} \Q(\f{i}{-k^2}\W)^n \tr \Q[(iB)^n\W]&=\f{i}{2n}\f{\Omega_3}{(2\pi)^4} \tr {B^n} \Q(\int_0^\Lambda dk\, \f{k^3}{k^{2n}}\W),
\een

where a Wick rotation was performed. The anomaly in Fujikawa's method was given in Eq. \eqref{11} as $\ma A=\sum \limits_{n=0}^\infty \f{(-1)^n}{2 n! }\f{\Omega_3}{(2\pi)^4} \Q(\text{tr}\, B^{n}\W)\Lambda^{4-2n}$. Following the renormalization group analysis of \cite{colewein}, we apply the operator $\f{\p}{\p \ln \Lambda}=\Lambda \f{\p}{\p \Lambda}$ to Eq. \eqref{3c}. Then from the fundamental theorem of calculus $\Lambda \f{\p}{\p \Lambda} \int^\Lambda_0\f{k^3}{k^{2n}}=\Lambda^{4-2n}$, we get the result that:  

\en \label{4c}
-\f{\p}{\p \ln \Lambda}V_{\text{eff}}=\sum \limits_{n=0}^\infty \f{1}{2n}\f{\Omega_3}{(2\pi)^4} \Q(\text{tr}\, {B^n} \W)\Lambda^{4-2n}.
\een

Only for $n=2$ does this match the anomaly given by Fujikawa's method. Indeed, it is impossible to construct a regulator in Fujikawa's method that exactly produces Eq. \eqref{4c}. However, the terms for $n \geq 3$ vanish in the limit $\Lambda \rightarrow \infty$. Since diagrams for which $n \geq 3$ are convergent, they do not contribute to the anomaly, and in Fujikawa's method they correspond to the vanishing $n \geq 3$ terms in the Taylor expansion. The anomaly is contained entirely in Fig. 2(b). The quadratic divergence in Fig. 2(a) is a well-known artifact of cutoff regularization and can be avoided by dimensional regularization, where the loop integral is zero \cite{mess}. However, Fujikawa's method does not work with dimensional regularization since in $d-2\epsilon$ dimensions, the $\delta$-function is zero \cite{brown}.\footnote{However, in the nonrelativistic context this need not be the case \cite{insert}.} Within the context of dimensional regularization, the anomaly arises from the fact that $\lambda \phi^4$ in $d-2\epsilon$ dimensions is not conformally invariant \cite{hoff} rather than through the noninvariance of the path integral measure.\\

This can readily be seen by calculating the effective potential. The effective potential is given by summing across all $n$ of Eq. \eqref{3c}:

\en \label{5c}
V_{\text{eff}}=-\sum\limits_{n=1}^\infty \f{1}{2n}\int \f{d^4k}{(2\pi)^4} \Q(\f{1}{k^2}\W)^n \tr B^n.
\een

One can swap the integral with the summation: this avoids the need for an IR regulator, as the summation results in a $\log $ which is IR-free. However, we are interested in the contribution of each n-vertex diagram -- therefore we introduce a fictitious mass $m$ to regulate the theory in the IR, and a cutoff $\Lambda$ to regulate the theory in the UV: 

\en \label{6c}
-V_{\text{eff}}&=\sum\limits_{n=1}^\infty \f{1}{2n}\int \f{d^4k}{(2\pi)^4} \Q(\f{1}{k^2+m^2}\W)^n \tr B^n \\
&=\f{1}{2}\int \f{d^4k}{(2\pi)^4} \Q(\f{1}{k^2+m^2}\W) \tr B+
\f{1}{4}\int \f{d^4k}{(2\pi)^4} \Q(\f{1}{k^2+m^2}\W)^2 \tr B^2+
\sum\limits_{n=3}^\infty \f{1}{2n}\int \f{d^4k}{(2\pi)^4} \Q(\f{1}{k^2+m^2}\W)^n \tr B^n.
\een
 
The integrals are standard, and the result in the $m^2 \rightarrow 0$ limit is:

\en \label{7c}
\f{1}{2}\int \f{d^4k}{(2\pi)^4} \Q(\f{1}{k^2+m^2}\W) \tr B&=-\f{\Lambda^2}{32 \pi^2}\tr B \\
\f{1}{4}\int \f{d^4k}{(2\pi)^4} \Q(\f{1}{k^2+m^2}\W)^2 \tr B^2&=\f{1}{64\pi^2}\Q[1-\log\Q(\Lambda^2/m^2 \W)\W] \tr B^2\\
\sum\limits_{n=3}^\infty \f{1}{2n}\int \f{d^4k}{(2\pi)^4} \Q(\f{1}{k^2+m^2}\W)^n \tr B^n &=\f{1}{128 \pi^2} \tr \Q[-3B^2+2B^2\log\Q(\f{-B}{m^2}\W)\W].
\een

One can see that diagrams with $n\geq 3$ are independent of $\Lambda$, and that $-\f{\p}{\p \ln \Lambda}$ acting on $n=2$ produces the anomaly. Both $\tr B=-\lambda(N+2)\phi_k\phi_k$ and $\tr B^2=\lambda^2(N+8)(\phi_k\phi_k)^2$ are of the form of the original Lagrangian, so can be cancelled by counter-terms. Adding all the terms in Eq. \eqref{7c} gives:

\en \label{8c}
V_{\text{eff}}=-\f{\Lambda^2}{32 \pi^2}\tr B -\f{\tr B^2}{128 \pi^2}+\f{1}{64 \pi^2}\tr \Q[B^2\log\Q(\f{-B}{\Lambda^2}\W)\W].
\een

The result is independent of $m^2$ as it should be. The $n \geq 3$ terms have produced a nonpolynomial $\log$ interaction, and the $n=2$ term has provided the scale for this interaction.

\section{Noether's Theorem and Dimensional Transmutation}

The field $\phi_c$ obeys the classical equations of motion Eq. \eqref{3b}, with the effective action $\Gamma[\phi_c]$ replacing the classical one $S[\phi_c]$. Therefore, Noether's theorem, which is based on the classical EOM, would apply if $\Gamma[\phi_c]$ retains the symmetry. In general the quantum corrections will create terms in $\Gamma[\phi_c]$ that explicitly break scale symmetry. The measure of symmetry-breaking is $\sum\limits_{i=1}^N \f{\p V_\text{eff}}{\p \phi_{ic}}\phi_{ic}-4V_\text{eff}$, which gives zero for the classically scale-invariant tree-level contribution $V=\f{\lam}{4} (\phi_{ic}\phi_{ic})^2$ to the effective potential. Specializing to $N=1$ the effective potential Eq. \eqref{8c} reads:

\en \label{2d}
V_\text{eff}=\f{\lambda \phi_c^4}{4}+\f{9 \lambda^2\phi_c^4}{64\pi^2}\Q(\ln \Q(\f{3\lambda \phi^2_c}{\Lambda^2}\W)-\f{1}{2} \W).
\een

Applying $\sum\limits_{i=1}^N \f{\p V_\text{eff}}{\p \phi_{ic}}\phi_{ic}-4V_\text{eff}$ to Eq. \eqref{2d}, we get the scale anomaly:

\en
\ma A=\f{9\lambda^2 \phi^4_c}{32 \pi^2},
\een

in agreement with Eq. \eqref{12}. From the viewpoint of classical physics, a term like $\phi_c^4 \ln M^2$ is scale-invariant, acting like a $\phi_c^4$ potential. It is $\phi_c^4 \ln \phi_c^2$ term that breaks scale-invariance. Both terms are related since dimensional transmutation of the $n=2$ graph provides the scale for the $n\geq 3$ graphs which generate nonpolynomial interactions.

\section{Conclusion}

The scale anomaly, and anomalies in general, are the result of the failure to maintain classical symmetry upon quantization. One cannot regularize the system in a way to preserve all the symmetries of the theory. The absence of dimensionful parameters in the action is sufficient for the classical theory to be scale invariant. However, the introduction of a dimensionful parameter through regularization can provide a scale to support non-invariant $\phi^{2n}$ interactions with $n\geq 3$ in the $O(N)$ quantum theory. Fujikawa's method is equivalent to the 1-loop calculation of the anomaly in the effective potential. \\

We plan to investigate these connections and apply the insights gained to the nonrelativistic case in order to study questions of interest in atomic and molecular physics, in particular in the field of ultracold atoms where, unlike the situation in particle physics, the manifestations of the scale anomaly in these systems have only now been accessible to experimentalists in this decade. 

\begin{acknowledgements}
This work was supported in part by the US Army Research Office Grant No. W911NF-15-1-0445.
\end{acknowledgements}


\end{document}